\documentclass[prl,floatfix,showpacs,amsmath,twocolumn]{revtex4}
\usepackage{amssymb}
\usepackage{graphicx}

\begin{document}

\newcommand{\be}{\begin{eqnarray}}
\newcommand{\ee}{\end{eqnarray}}
\newcommand{\bea}{\begin{eqnarray}}
\newcommand{\eea}{\end{eqnarray}}
\newcommand{\bma}{\begin{subequations}}
\newcommand{\ema}{\end{subequations}}
\newtheorem{lemma}{Protocol}
\def\ket #1{\vert #1\rangle}
\def\bra #1{\langle #1\vert}

\def\lR{l^2_{\mathbb{R}}}
\def\RR{\mathbb{R}}
\def\E{\mathbf e}
\def\D{\boldsymbol \delta}
\def\S{{\cal S}}
\def\T{{\cal T}}
\def\dd{\delta}
\def\one{{\bf 1}}
\def\Flip{{\cal F}}

\title{Delocalized Entanglement of Atoms in optical Lattices}

\author{K. G. H. Vollbrecht$^1$ and J. I. Cirac$^1$}
\affiliation{$1$ Max-Planck Institut f\"ur Quantenoptik,
Hans-Kopfermann-Str. 1, Garching, D-85748, Germany }

\date{\today}

\begin{abstract}
We show how to detect and quantify entanglement of atoms in optical lattices in terms of correlations functions of the
momentum  distribution. These distributions can be measured directly in the experiments.
We introduce two kinds of entanglement measures related to the position and the spin of the atoms.

\end{abstract}

\maketitle
Experiments on atoms in optical lattices have recently become very
attractive playgrounds to investigate basic issues in the
context of quantum information theory \cite{1C,2C,3C}. The high degree of control
reached in those experiments should allow us to prepare a large
variety of entangled states, to analyze their physical properties,
and to verify and quantify the presence of entanglement. However,
in order to carry out these investigations, a clear definition
of entanglement with a clear physical meaning should be given,
and ways to detect it should be explored \cite{tot}. For atoms in optical
lattices those questions are far from being trivial, since one has to
consider different degrees of freedom (such as atom numbers and internal
levels) and take into account the presence of super-selection rules,
as well as the fact that in most of those experiments we only have
access to certain collective properties.

In this letter we will define and explore certain kinds of entanglement
which are relevant in current experiments dealing with bosonic and
fermionic atoms in optical lattices. We will concentrate on the
entanglement properties between different sites of the optical lattice
(i.e. in second quantization), since in this case a clear meaning
as a resource for quantum information tasks can be assigned to
those definitions. Furthermore, we will restrict ourselves here to
the simplest case of bipartite entanglement. Since it is very hard
in practice to address atoms at different lattice sites, we  define the
{\em delocalized} bipartite reduced density operators for a state
 $\rho$ on  an optical lattice by
\bea\label{statedef}
\rho_{AB}=\sum_m \rho_{(m,m+x)},
\eea
where $\rho_{(m,m+x)}$ denotes the restriction of $\rho$ to the
sites $m$ and $m+x$ of the lattice. Since we do not want to rely on
any form of addressability, we assume the lattice to be (approximately) infinite.
To simplify matters, we consider here only
a 1D lattice, but everything  holds as well when the state is defined on 2D or 3D lattices.
We  will study the entanglement of $\rho_{AB}$ by
means of experimental feasible collective measurements, that   can be translated directly into expectation values
of $\rho_{AB}$.
Note that, for this one to one correspondence between  measurement
results and expectation values of $\rho_{AB}$, we define this state to be unnormalized.
This definition  of $\rho_{AB}$ is useful for entanglement detection, however for a quantitative investigation, we need to
include the normalization in the definition.
Due to the infinite number of lattice sites the trace of $\rho_{AB}$ is infinite, such
that a straight forward normalization fails to give any utilizable quantitative information.
Therefore we have to take into account that, for a finite number of atoms, most of the lattice sites are
empty and do not contribute to any measurement.
The obvious solution is to restrict the summation  to a finite part of the lattice, such that $\rho_{AB}$ can be normalized.
But this makes only sense if we can ensure the atoms to be located in a relative small region of the lattice, what requires
either some form of addressability or an extra assumption about the localization of the state $\rho$.
A more general solution is to define the state $\rho'_{AB}$ as the (normalized) projection of $\rho_{AB}$
to the subspace where at least one atom is present, i.e., project out the zero atom subspace of $\rho_{AB}$. Even if the atoms are
evenly spread over  the whole lattice,
as long as the number of atoms is small, the projection of $\rho_{AB}$ has a finite trace.
Unfortunately,  we can not give the exact trace in terms of experimental feasible measurements, but we can bound the trace
which still allows us to derive lower bounds for the entanglement.
Through the  definition of $\rho_{AB}$ and $\rho'_{AB}$ we are mixing different contributions of different sites, which
implies that the entanglement we define will be somehow delocalized
between different pairs of sites which are separated by a distance $x$. Note, that
 $\rho_{AB}$ depends on the distance $x$, which can be freely chosen.
We will give lower bounds to the {\it entanglement of formation} for $\rho'_{AB}$.
 Note, that due to the translationally symmetry of (\ref{statedef}), we can never reach a maximally entangled
state for $\rho'_{AB}$. {The maximally possible entanglement of formation that can be reached is $0.285$. }\cite{Woot}

We will consider two situations: firstly,
that in which the atomic internal levels are not involved, and
therefore we will deal with different occupation numbers; secondly,
the one in which entanglement occurs between different internal
states of the atoms in each site. As we will show, the first
case is very simple to characterize and one can already claim
that this kind of entanglement has been created in several
experiments carried out so far. The second kind of entanglement
is much subtler, and require more sophisticated measurements
in order to prove the existence of entanglement.

The trapped atoms are described in second quantization by the annihilation and creation operators.
We will assume two internal levels $a$ and $b$ for each atom, in which
$a_m,a^\dagger_m$ resp. $b_m, b^\dagger_m$ denote the  annihilation and creation operators of
atoms in level $a$ resp. $b$ at site $m$.

{\it Measurements:}
One kind of measurement that is feasible in  experiments is to turn off
the lattice potential and look at the  density $n_a(x,t)=\left< \psi_a(x,t)^\dagger\psi_a(x,t)\right>$ (resp. $n_b(x,t)$) of the expanding
atom cloud, where $\psi_a(x,t)$ is the bosonic or fermionic field operator for the internal level $a$.
Furthermore, one can measure the density-density correlations  $c_{ab}(x,x',t)=\left< \psi_a(x,t)^\dagger\psi_a(x,t)\psi_b^\dagger(x',t)\psi_b(x',t)\right>$
and in a similar  way $c_{aa}(x,x',t),c_{bb}(x,x',t) $ and $c_{ba}(x,x',t)$  \cite{xyz}.
In a long time of flight approximation the new
positions of the atoms can be detected and becomes proportional to the initial momentum distribution $k\approx m x/(\hbar t)$.
Due to this relation
we can measure the momentum distribution   via
${n_a}(k)= \lim_{t\rightarrow\infty}n_a(\frac{\hbar k t}{m},t)$ and their correlations $c_{xy}(k,k')= \lim_{t\rightarrow\infty} c_{xy}(\frac{\hbar k t}{m},\frac{\hbar k' t}{m},t)$.
This momentum distribution   in second quantization is given by
\bea \label{na}
{n_a}(k)\approx \sum_{n,m}\hat{w}_n(k) \hat{w}_m(k)^* \left<a_n^\dagger \hat{a}_m \right>
\eea
and the density-density correlations by
\bea\label{cab}
 &&c_{ab}(k,k')\approx \\ \nonumber
 &&\sum_{n,m,n',m'}\hat{w}_{n}(k)\hat{w}_{m}(k)^*\hat{w}_{n'}(k')\hat{w}_{m'}(k')^* \left<a_n^\dagger a_m b_{n'}^\dagger b_{m'}\right>,
 \eea
where $\hat{w}_n(k)$ denotes the
Fourier transformed Wannier function at time zero at site $n$.
 And in an analogous manner we get $n_b(k)$,  $c_{aa}(k,k'),c_{bb}(k,k')$ and $c_{ba}(k,k')$.

{\it Occupation  number entanglement:}
We assume now the simple case, where we  only have one type
of atom, say that in level $a$. Due to
the atom number conservation the possible product states
in second quantization are restricted to be of the form
\bea \label{basis}
\ket{\dots ,n_{0},n_1,n_2, \dots  },
\eea
because it is not allowed to have superpositions between
states with different number of atoms. Here $n_i$ denotes the occupation number
of sites $i$. We define a state to be entangled, if it
can not decomposed into states of the form (\ref{basis}).
Testing  separability simplifies in this case to check whether a given state
is diagonal in (\ref{basis}). Note, that entanglement defined with respect to such a super-selection
rule is in general a less powerful resource for quantum information tasks, because the entanglement properties
can only be seen/used when having access to several copies of the state \cite{Norbert}.
To detect this kind of entanglement we look at the momentum distribution (\ref{na}).
A simple observation is that for all product states we get a  $\delta(n=m)$ in the sum, because
$a_n^\dagger a_m$ maps any state of form ($\ref{basis}$) to a orthogonal one if $n\neq m$. For
this reason we get for any
separable state that $n_a(k)=N$ is just proportional to the total number operator  and
is independent from $k$, i.e., the momentum distribution is flat (up to the envelope Wannier-functions).
Any non flat momentum distribution indicates an entangled
multipartite state, something that has been already observed in several experiments \cite{3C}.
For a more  quantitative statement about the entanglement of (\ref{statedef}) we look at the Fourier transformation of the momentum distribution of $\rho$
\bea\label{2}
\left<Q_x \right>_{\rho}:=\int_{} dk \ e^{-ik x d} \left< n_a(k)\right>_\rho= \sum_m \left<a_m^\dagger a_{m+x}\right>_\rho.
\eea
Here we have used, that multiplication by a phase $e^{ikxd}$ in momentum representation results in a shift in position, i.e.,  $\hat{w}_n(k) e^{-ik xd}=\hat{w}_{n+x}(k)$
and that two at different places located Wannier functions are orthogonal, i.e.,  $\int dk \ \hat{w}_{n+x} \hat{w}_{m}^*=\delta(n+x,m)$.
$x$ is taken to be an arbitrarily integer and $d$ denotes the lattice constant.
For (\ref{2}) we can give an interpretation in terms of an expectation value of the bipartite
state $\rho_{AB}$ (\ref{statedef}):
\bea\label{wert1}
\left<Q_x \right>_{\rho}:=\left<a_A^\dagger a^{}_B\right>_{\rho_{AB}}.
\eea
Assume now the idealized case where the occupation number of every site is restricted to be either one or zero, defining this way
exactly one qubit per site.
Then $\rho_{AB}$ is a two qubit density matrix and $\left<a_A^\dagger a^{}_B\right>=\bra{01}\rho_{AB}\ket{10}$ is exactly
one off-diagonal element, where $\ket{0}$ and $\ket{1}$ denote  empty or  occupied sites.
Due to the super-selection rules it is the only allowed off-diagonal element and defines the entanglement
properties of the state.
The state $\rho_{AB}$ can be decomposed into two parts.
A separable part spanned by  the vectors $\ket{11},\ket{00}$ and
the part
spanned by the vectors $\ket{10},\ket{01}$ that contains entanglement if
$\bra{01}\rho_{AB}\ket{10}\neq 0$. For this part we now want to estimate a lower bound
for {\it entanglement of formation} \cite{eof}.
Note, that we use here  a definition for  entanglement of formation  respecting the super-selection rules.
Assuming now a normalized state $\rho'_{AB}$ with given off-diagonal element $\lambda$,
it can easily be shown, that a pure state completely supported on the $\ket{10},\ket{01}$ subspace having
the same off-diagonal element $\lambda$ gives
us a lower bound to the entanglement of formation. Exploiting this
leads to a lower bound for the entanglement  given by
$
E_{of}(\rho'_{AB})\geq S\left(\frac{1}{2} \left[1-\sqrt{ 1-4{|\lambda|^2}}\right]\right),
$
where $S(x)=-x \log(x)-(1-x) \log(1-x)$ denotes
the  von Neumann entropy. 
To estimate $|\lambda|$ for our $\rho'_{AB}$ we first have to find a  bound for the trace of the unnormelized $\rho'_{AB}$.
This can be given by $2\left<N\right>$, since the the reduced densities sates in (\ref{statedef}) cover two times the whole lattice.
Therefore we can conclude that $|\lambda| \geq \frac{\left<Q_x \right>_{\rho}}{2\left<N\right>}$.

While the assumed restriction of maximal one atom per site matches perfectly for fermions,
in the bosonic case we can still give a bound if the following constraint can be guarantied, e.g.,  verified
by further measurements \cite{CirZo}. The expected number of sites with more than one atom has to be smaller than $\epsilon \left<N\right>$
and the maximally occupation number of one site  has to be smaller than $r$.
Under these conditions, the measurement result is still close to the off-diagonal element $\bra{01}\rho_{AB}\ket{10}$.
The error coming from overpopulated sites can bounded  by $(2\epsilon r+4\sqrt{ \epsilon}) \left<N\right> $ (see Appendix)
leading to
\bea
&&E_{of}(\rho'_{AB})\geq \\
&&S\left(\frac{1}{2} \left[1-\sqrt{ 1-\frac{(|\left<Q_x\right>|-(2\epsilon r+4\sqrt{\epsilon}) \left< N \right> r)^2}{\left< N \right>^2}}\right]\right).\nonumber
\eea

{\it Internal level entanglement:}
We now consider the case, where we have two level atoms in the lattice.
In the ideal situation we would have exactly one atom per site such that the internal levels realize one
qubit. In this case $\rho_{AB}$ is again a two qubit state without any restriction due
to the conservation laws.
To detect this stronger kind of entanglement it is not enough to look independently at the
momentum distribution of level $a$ and $b$, but we have to look
at the correlation \cite{xyz} between  momentum distributions $c_{aa}(k,k'), c_{bb}(k,k'),c_{ab}(k,k')$ and $c_{ba}(k,k')$.
By properly chosen Fourier transformations in $k$ and $k'$ we define
\bea
\left<Q_x^{ab}\right>&:=&\int \int dk \ dk' \ e^{i k x d} e^{-ik' x d} c_{ab}(k,k')\\
&=& \sum_{m  m' }  \left< a_m^\dagger a_{m+x}^{} b_{m'+x}^\dagger b_{m'}^{} \right>,\label{4}
\ee
and in an analog way $\left<Q_x^{aa}\right>$ and $\left<Q_x^{bb}\right>$. Here we again use
the fact that the integrals over Wannier functions on different sites leads to delta functions.
We furthermore assume a situation where we  can restrict these sums to the case
$m=m'$.
\bea
\left<Q_x^{ab}\right>= \sum_{m}  \left< a_m^\dagger a_{m+x}^{} b_{m+x}^\dagger b_{m}^{} \right> \label{Qxab}
\eea
We will discuss later in the section\textit{ dephasing} how this condition can be realized by adding extra magnetic fields
such that the $m \neq m'$ terms vanish.

{\it  The one atom per site case:}
To illustrate the basic idea, we
assume now the idealized situation, where we have a state $\rho$
for which we can ensure, that in every site
is exactly one atom. 
Note, that in this case we can use (\ref{Qxab})  without assuming any extra magnetic fields,
because the terms with $m\neq m'$ vanish already because of the assumption.
Equation (\ref{Qxab}) can now interpreted as the expectation value
\bea\label{Qab}
\left<Q_x^{ab}\right>_{\rho} :=\left< a_A^\dagger a_{B}^{} b_{B}^\dagger b_{A}^{} \right>_{\rho_{AB}}
\eea
 of a  bipartite density matrix $\rho_{AB}$ as defined in (\ref{statedef}). We want now
 to calculate the overlap of the state $\rho'_{AB}$ with a maximally entangled state, i.e.,
 the fidelity $f_{\Phi}(\rho'_{AB})=\bra{\phi} \rho'_{AB}\ket{\phi}$, where $\phi$ will be one of the Bell-state
 defined by
\bea
\phi_\pm=\frac{1}{\sqrt{2}}( \ket{10}_A\ket{01}_B\pm \ket{01}_A\ket{10}_B ) \nonumber\\
\psi_\pm=\frac{1}{\sqrt{2}}( \ket{10}_A\ket{10}_B\pm \ket{01}_A\ket{01}_B ) \nonumber
\eea
Here $\ket{10}_A$
denotes the atom of Alice being in the $a$ and $\ket{01}_A$ being in the $b$ level and in analog way for Bob.
We claim that now, that
\bea
 \left<Q_x^{ab}+Q_x^{ba} \right>_\rho&=&-\left<\phi_-\right>_{\rho_{AB}}+\left<\phi_+\right>_{\rho_{AB}} \nonumber \\
 \left<Q_x^{aa}+Q_x^{bb} \right>_\rho&=&\pm\left(\left<\psi_-\right>_{\rho_{AB}}+\left<\psi_+\right>_{\rho_{AB}}\right)+\left<N\right>_\rho, \label{pm}
\eea
where the $(\pm)$ in (\ref{pm}) distinguishes between the bosonic and the fermionic case.
This can easily checked, by calculating the expectation values  for an arbitrary pure state
\bea \ket{\Phi}&=&\lambda_{00}\ket{10}_A\ket{10}_B+\lambda_{01}\ket{10}_A\ket{01}_B\\ \nonumber
       &&+\lambda_{10}\ket{01}_A\ket{10}_B+\lambda_{11}\ket{01}_A\ket{01}_B.
\eea
We get that $\left<Q_x^{ab}\right>_{\Phi}=\lambda_{01}\lambda_{10}^*$  and $\left<Q_x^{ba}\right>_{\Phi}=\lambda_{01}^*\lambda_{10}$
such that the sum is equal to
 $\left<-\ket{\phi_-}\bra{\phi_-}+\ket{\phi_+}\bra{\phi_+}\right>_{\Phi}$.
Furthermore we obtain that
\bea \left<Q_x^{aa}\right>_{\rho} :&=&\left< a_A^\dagger a_{B}^{} a_{B}^\dagger a_{A}^{} \right>_{\rho_{AB}}\\ \nonumber
&=&\left<  \pm a_A^\dagger a_{A}^{}   a_{B}^\dagger a_{B}^{}+a_A^\dagger a_{A}^{}  \right>_{\rho_{AB}}, \label{fermi}
\eea
where the $(\pm)$ distinguishes the bosonic from the fermionic case and in analogous manner for $\left<Q_x^{bb}\right>$. It is
easily verified, that
$\left< a_A^\dagger a_{A}^{}   a_{B}^\dagger a_{B}^{}\right>+\left<   a_A^\dagger a_{A}^{}   a_{B}^\dagger a_{B}^{}\right>=
|\lambda_{00}|^2+|\lambda_{11}|^2=\left<\psi_-\right>+\left<\psi_+\right>$ and $\left<a_A^\dagger a_{A}^{}+b_A^\dagger b_{A}^{}\right>=\left<N\right>$.$\square$

From (\ref{pm}) it is now easy to derive the fidelities $f_{\phi_\pm}$  for $\rho'_{AB}$.
Due to the one atom per site assumption we can bound the trace of the projected $\rho_{AB}$ by $\left<N\right>$
(instead of 2$\left<N\right>$)
leading to
\bea
&&f_{\phi_\pm}(\rho'^B_{AB})\geq\frac{1}{2}
+\frac{\pm \left<Q^{ab}+Q^{ba}\right>_{\rho}+\left<N-Q^{aa}-Q^{bb}\right>_{\rho}}{2\left<N\right>} \nonumber\\
&&f_{\phi_\pm}(\rho'^F_{AB})\geq
\frac{1}{2}
+\frac{\pm \left<Q^{ab}+Q^{ba}\right>_{\rho}+\left<Q^{aa}+Q^{bb}-N\right>_{\rho}}{2\left<N\right>},\nonumber
\eea
for bosonic and fermionic case.
Note, that we can get fidelities with further maximally entangled states
by applying a global unitary $U \otimes \dots \otimes U$ to $\rho$. This translates to apply $U \otimes U$ to $\rho_{AB}$. While
the overlap with $\ket{\phi_-}$ stays unchanged,  we get that
$$\bra{\phi_+}U \otimes U \rho_{AB} (U\otimes U)^\dagger\ket{\phi_+}=\bra{\phi_U} \rho_{AB} \ket{\phi_U},$$
where $\ket{\phi_U}=\one \otimes (UU^T)\ket{\phi_+}$ is a maximally entangled state.
In particular, we can get  $f_{\psi_\pm}$ by properly chosen $U$.
The fidelities $f_{\phi_\pm}$ and $f_{\psi_\pm}$
of a state   directly gives us lower bounds to several entanglement measurements by comparing
it to Bell-diagonal or Isotropic states \cite{eof}, e.g.,
\bea\label{ergeof}
E_{of}(\rho'_{AB})\geq S\left(\frac{1}{2}\left[1-\sqrt{1-\left(1-2f_{\phi_\pm}\right)^2}\right]\right).
\eea

{\it The general case:}
We consider now the general case where the number of atoms per site is arbitrarily.
In this case is it difficult to give a lower bound to $\bra{\phi_\pm}\rho_{AB}\ket{\phi_\pm}$, because
the four Bell-states do not build a basis in the larger Hilbert-space and it is  therefore not possible
to derive the fidelity
from (\ref{pm}). But it still makes sense to define
$\Lambda:= \pm (\left<\phi_+\right>-\left<\phi_-\right>)+\left<\psi_+\right>+\left<\psi_+\right>$ in the larger Hilbert-space.
Given a value of $\Lambda$ it can be shown that the bound (\ref{ergeof}) still holds with
$f_{\phi_\pm}$ now replaced by $f_{\phi_\pm}=\frac{1-\Lambda}{2}$, even though
 $f_{\phi_\pm}$ has now no direct interpretation as  fidelity.
 Note, that $\Lambda$ is a direct bound for the concurrence \cite{eof} and can be used itself to quantify the entanglement.
Our goal is to give a lower bound for $f_{\phi_\pm}$ for a state $\rho''_{AB}$ that we define
in this case as the projection of $\rho_{AB}$ to the subspace with $2$ or more atoms. We assume, that
number of  defective sites $D$ should be bounded by  $D \leq \epsilon \left< N\right>$.
As defect counts every site having two or more atoms. Furthermore we assume the maximal occupation number of $a$ and $b$ to be less than $r$.
A straight forward calculation (see Appendix) leads to
\bea
&&f^B_{\phi_\pm}\geq
\frac {\left<\pm(Q^{ab}+Q^{ba})+(2-4\epsilon r^2)N-Q^{aa}-Q^{bb}\right>_\rho}{2\left<N\right>_\rho}\nonumber\\
&&f^F_{\phi_\pm}\geq
\frac {\left<\pm(Q^{ab}+Q^{ba})+Q^{aa}+Q^{bb}-4\epsilon N\right>_\rho}{2\left<N\right>_\rho}\nonumber
\eea
for the bosonic and fermionic cases, which can be used to estimate a lower bound  for $\rho''_{AB}$.

{\it Dephasing with magnetic field:}
We  discuss now the possibilities of eliminating  the terms of (\ref{4}) where $m \neq m'$
by dephasing, i.e., by destroying any delocalization of atoms over several sites.
 If we write $\Phi=\sum_K \lambda_K \ket{K}$ in a product basis $\ket{K}=\ket{k^a_{-\infty},\dots k^a_\infty, k^b_{-\infty},\dots k^b_\infty }$,
 where $k^{(a/b)}_i$ denotes the number of atoms in site $i$ in level $a/b$,
then the unwanted terms that contribute to the sum are of the form
\bea \label{mist}
\lambda_K^* \lambda_{K'}^{}\left<K\right|  a_m^\dagger a_{m+x}^{} b_{m'+x}^\dagger b_{m'}^{}  \left| K'\right>,
\eea
where $m \neq m'$.
To give a nonzero
value,
 $\ket{K}$ has to be equal to $\ket{K'}$ although $\ket{K}$ has an extra atom
in $m$ and $m'+x$ and a missing atom in $m+x$ and $m'$, which implies defects in this pair of sites.
We will use this displaced atoms
and an additional  inhomogeneous magnetic field to introduce some random phase to (\ref{mist}) and
therewith guarantee that these terms  vanish.
Let us assume a magnetic field, that is proportional to $k^2$, where $k$ is the number of the lattice
site. Then the state $\ket{K'}$ gets, up to a global phase,  a phase of $e^{i(m+x)^2 t+im'^2t}$, whereas
$\ket{K'}$ picks up $e^{i m^2t+i(m'+x)^2t}$. So $\ket{K}$ and $\ket{K'}$ get a relative phase
of $e^{i 2(m'-m)x t}$ \footnote{We assumed, that atoms in level $a$ and $b$ gets the same phase independent of the internal state.}.
 By a properly chosen set of times, we can randomize the phases such
that for given $x$ all terms with $m\neq m'$ vanish in average, while the terms with $m=m'$ stay unchanged.
Note that even without magnetic field it seems to be quite unlikely,
that all terms of the form (\ref{mist}) sum up to a nonzero contribution.
To give some non vanishing amount, the defects have to be correlated in a very
unlikely way.
In detail, one defect located at $m$ or $m+x$
has to be correlated to a defect located at $m'$ or $m'+x$ and in addition all these cases has to be correlated with each other.
If we assume, that defects occur only randomly
and therefore are uncorrelated, (\ref{mist}) already vanish.

In conclusion, we have defined two figures of merit to quantify entanglement for atoms trapped in optical
lattices, by only measuring density correlation functions of the expanded atomic cloud, without requiring
any addressability of the original lattice setup. One set of measurement data can be used to study
entanglement at arbitrarily distances $x$.
We discuss bounds in the cases where the defects in
the lattice can be bounded. We acknowledges support from EU projects SCALA
and DFG-Forschungsgruppe 635.

\subsection{Appendix}
{\it Calculating errors for the off-diagonal element:}
$\rho_{AB}$ can be written as a direct sum $\sum_n \rho_n$, where $\rho_n$ is the n-atom subspace.
The off-diagonal element $\bra{01}\rho_{AB}\ket{10}$ is exactly given by $\left<Q_x\right>_{\rho_1}$.
So we have to bound all the absolute values of $\left<Q_x\right>_{\rho_n}$ for $n>1$. Note, that all $\rho_n$ for $n>2$ has at least
one defect. Therefore we can bound the trace by two times the number of defects ($2 \epsilon \left<N\right>$)
and therefore the total contribution by  $2 \epsilon \left<N\right> r$.
We need to take twice the number of defects, because the reduced states in (\ref{statedef}) cover two times the whole lattice.
More complicated is to find a bound $\rho_2$, because 2 atoms do not imply automatically a defect. We therefore
look at a pure state supported on $\rho_2$
\bea
\ket{\psi_2}=\lambda_{0}\ket{11}+\lambda_1\ket{20}+\lambda_{2}\ket{02}.
\eea
$\left<Q_x\right>_{\psi_2}$ easily calculates to $(\lambda_{0}\lambda_1^*+\lambda_{0}^*\lambda_2)\sqrt{2}$,
with $\sum_i |\lambda_i|^2=1$.
We now try to find an upper bound for  its absolute value. First lets assume, that all
$\lambda_i$ are real and positive leading to $\lambda_{0}(\lambda_1+\lambda_2)\sqrt{2}$. For given
$\lambda_0$ this is maximized for $\lambda_1=\lambda_2$. Defining $\lambda'=\frac{1}{\sqrt{2}}\sqrt{\lambda_1^2+\lambda_2^2}$,
we get $\left<Q_x\right>_{\psi_2} \leq\lambda_{0}2 \lambda' \sqrt{2}\leq2 \lambda' \sqrt{2} \leq 2\sqrt{\lambda_1^2+\lambda_2^2}$ as upper bound.
Now, we use that $\sqrt{x}\leq  \frac{1}{2\sqrt{R}}x+\frac{\sqrt{R}}{2}$ for any positive parameter $R$, which gives us
$\left<Q_x\right>_{\psi_2}\leq \frac{1}{\sqrt{R}}(\lambda_1^2+\lambda_2^2)+\sqrt{R}$.
Here  $(\lambda_1^2+\lambda_2^2)$ is now exactly the probability for an defect. For the unnormelized $\rho_2=\sum_i q_i \ket{\psi_2^i}\bra{\psi_2^i}$
we get
$\left<Q_x\right>_{\rho_2}\leq \frac{1}{\sqrt{R}}\sum_i q_i (\lambda_1^2+\lambda_2^2)_i+\sqrt{R} \ tr(\rho_2)$.
The sum $\sum_i q_i (\lambda_1^2+\lambda_2^2)_i$ is now the probability to found a defect in $\rho_2$, that can be
bounded by $2 \epsilon \left<N\right>$. Furthermore, the trace of $\rho_2$ can be bounded by $tr(\rho_2)\leq \left<N\right>$, leading to
$\left<Q_x\right>_{\rho_2}\leq (\frac{1}{\sqrt{R}}\epsilon +\sqrt{R}) 2 \left<N\right>$. Now we a free
to choose $R=\sqrt{\epsilon}$  and get finally
$\left<Q_x\right>_{\rho_2}\leq  4 \sqrt{\epsilon} \left<N\right>$.
So we get
$|\bra{01}\rho_{AB}\ket{10}|= |\left<Q_x\right>_{\rho_1}| \geq|\left<Q_x\right>_\rho|- (4 \sqrt{\epsilon}+2 \epsilon r) \left<N\right>$

{\it Calculating errors for fidelity:}
We want to bound
$\Lambda=2f_{\phi_\pm}-1= \pm \left<\phi_+\phi_+\right>+\left<\psi_+\psi_+\right>$.
Given the state  $\rho_{AB}=\sum_n \rho_n$  written as direct sum.
We are now interested in
$\Lambda_{\rho_{AB}}$ which is equal to $\Lambda_{\rho_2}$,
because $\Lambda$ is supported on the 2-atom subspace.
Since (\ref{pm}) holds also on the bigger 2-atom subspace
$2f_{\phi_\pm}(\rho_2)-1$ equals $\Lambda_{\rho_2}$.
$\Lambda_{\rho_1}$ gives no contribution and can therefore neglected. For this
reason, we only have to take defects with two or more atoms per site into account.
We therefore have only to bound  $2f_{\phi_\pm} (\rho^\bot)-1$, where  $\rho^\bot=\sum_{n>2}\rho_n$.
Note, that this sum runs by assumption only up to $r^4$.
The trace of $\rho^\bot$ is smaller than  $2 D<2 \epsilon\left<N\right>$, since for $n>2$ there exists always a defect and $2f_{\phi_\pm}^B-1 $ can be bounded
by $4r^2$ for bosons and by $4$ for fermions.

\end{document}